\documentclass[Physsubmission, Phys]{SciPost}
\hypersetup{
    colorlinks,
    linkcolor={red!50!black},
    citecolor={blue!50!black},
    urlcolor={blue!80!black}
}

\usepackage[bitstream-charter]{mathdesign}
\DeclareSymbolFont{usualmathcal}{OMS}{cmsy}{m}{n}
\DeclareSymbolFontAlphabet{\mathcal}{usualmathcal}

\begin{document}

\begin{center}{\Large \textbf{
Production of three isolated photons in the high-energy factorization approach\\
}}\end{center}

\begin{center}
Vladimir A. Saleev\textsuperscript{1$\star$}
\end{center}

\begin{center}
{\bf 1} Samara National Research University, Samara, Russia
\\
* saleev@samsu.ru
\end{center}

\begin{center}
\today
\end{center}


\definecolor{palegray}{gray}{0.95}
\begin{center}
\colorbox{palegray}{
  \begin{tabular}{rr}
  \begin{minipage}{0.1\textwidth}
    \includegraphics[width=22mm]{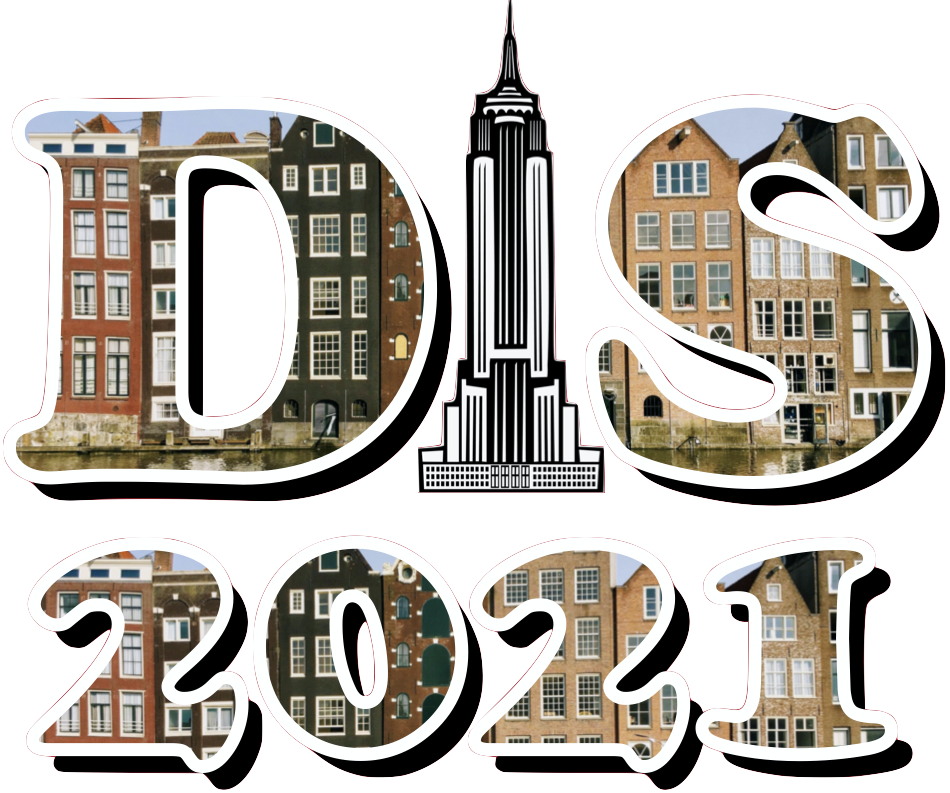}
  \end{minipage}
  &
  \begin{minipage}{0.75\textwidth}
    \begin{center}
    {\it Proceedings for the XXVIII International Workshop\\ on Deep-Inelastic Scattering and
Related Subjects,}\\
    {\it Stony Brook University, New York, USA, 12-16 April 2021} \\
    \doi{10.21468/SciPostPhysProc.?}\\
    \end{center}
  \end{minipage}
\end{tabular}
}
\end{center}

\section*{Abstract}
{\bf
We study large-$p_T$ three-photon production at the LHC at the
center-of-mass energy $\sqrt{s}=8$ TeV. We use the LO approximation
of the parton Reggeization approach consistently merged with the
real  NLO corrections. For numerical calculations use the
parton-level generator KaTie and modified KMR-type unintegrated
parton distribution functions. We find good agreement between our
predictions and data with the same accuracy as in the NNLO
calculations based on the collinear parton model of QCD. At higher
energies ($\sqrt{s}=13$ and 27 TeV) parton Reggeization approach
predicts larger cross sections, up to $\sim 10$ \% and $\sim 20$ \%,
respectively. }


\section{Introduction}
\label{sec:intro}
The recent experimental data for large-$p_T$ three-photon production
at the energy 8 TeV \cite{LHC8}  are extensively studied in the
collinear parton model (CPM) within framework of perturbative
approach of QCD beyond the leading-order (LO) accuracy in
strong-coupling constant $\alpha_S$, i.e. at the
next-to-leading-order (NLO) \cite {NLO31,NLO32} and even at
next-to-next-to-leading-order (NNLO) \cite{NNLO31,NNLO32}. The
high-order calculations for the three-photon production in CPM of
QCD provide rather bad agreement with data at the level of NLO
accuracy. Inclusion of the NNLO QCD corrections \cite{NNLO31,NNLO32}
eliminates the existing discrepancy with respect to NLO QCD
predictions. However, for three-photon production the agreement with
data is not so good as for single or two-photon production and it is
achieved when hard scale parameter $\mu$ is taken very small
\cite{NNLO31,NNLO32}.

In CPM we neglect the transverse momenta of initial-state partons in
hard-scattering amplitude that is correct assumption for the fully
inclusive observables,  such as $p_T$ spectra of single prompt
photons or jets, where their large transverse momentum defines
single hard scale of the process, $\mu \sim p_T$. The multi-photon
large-$p_T$ production is multi-scale hard process in which use the
simple collinear picture of initial state radiation may be a bad
approximation. In the present paper, we calculate different
multi-scale variables in three-photon production from a point of
view of high-energy factorization (HEF)
\cite{KTfactorization1,KTfactorization2}. We use the parton
Reggeization approach (PRA) which is a version of HEF formalism,
based on the  modified multi-Regge kinematics (mMRK) approximation
for QCD scattering amplitudes \cite{NSS2013,NKS2017}. This
approximation is accurate both in the collinear limit, which drives
the transverse-momentum-dependent (TMD) factorization and in the
high-energy (multi-Regge) limit, $\hat{s}\gg (-\hat{t})\sim {\bf
p}_T^2\sim \mu^2$.

In same manner of PRA, we studied previously one-photon production
~\cite{one-photon}, two-photon production \cite{two-photon} and
photon plus jet production ~\cite{photon-jet} in proton-(anti)proton
collisions at the Tevatron and LHC.

\begin{table} \label{Tab:1}
\begin{center}
\begin{tabular}{|c|c|c|} \hline
Hard scale, $\mu$ & $\sigma_{\mbox{LO}}$ [fb]  &
$\sigma_{\mbox{NLO}}$ [fb] \\ \hline
 $M_{3\gamma}/2$ & $31.07^{+8.87}_{-6.76} $ &
 $69.22^{+4.05}_{-1.07}$ \\ \hline
$p_{T,3\gamma}/2$ & $29.72^{+9.22}_{-6.72}$   &
$69.76^{+4.29}_{-1.85}$    \\ \hline
 $E_{T,3\gamma}/2$  &
$32.50^{+9.80}_{-2.65}$  & $71.00^{+4.93}_{-2.65}$  \\ \hline
\end{tabular}
\end{center}
\caption{Predictions for $p+p\to \gamma\gamma\gamma + X$ total cross
section at $\sqrt{s}=8$ TeV for the different choice of
factorization/renormalization scale ($\mu=\mu_F=\mu_R$), errors
indicate variation by factor two around the middle values which are
listed in first column.}
\end{table}

\begin{table}
\begin{center}
\begin{tabular}{|c|c|c|c|c|} \hline
$\sqrt{s}$[TeV] & $\sigma_{\mbox{\small LO}}$ &
$\sigma_{\mbox{\small NLO}}$ & $K_{\mbox{\small NLO}}$ &
$\sigma_{\mbox{\small NNLO}}^{\mbox{\small CPM}}$ \cite{NNLO32}
\\ \hline
8  & $32.50^{+9.80}_{-7.46}$   & $71.00^{+4.93}_{-2.65}$    &  2.18 & $67.42^{+7.41}_{-5.73}$\\
\hline
 13 & $53.91^{+18.14}_{-14.11}$ & $126.79^{+10.43}_{-7.30}$  & 2.35 & $114^{+13.64}_{-10.54}$\\ \hline
27 &  $115.25^{+45.09}_{-34.45}$ & $298.54^{+30.71}_{-25.55}$   & 2.59  & $245.91^{+32.46}_{-24.34}$\\
\hline
\end{tabular}
\end{center}
\caption{Predictions for $p+p\to \gamma\gamma\gamma + X$ total cross
section at the different center-of-mass energies, $\sqrt{s}$.
Numerical error of calculation
 is equal $0.1 \%$.}
\end{table}

\section{Parton Reggeization Approach}
\label{sec:PRA}
\subsection{High-energy factorization}
The cornerstones of PRA are $k_T-$dependent factorization formula,
unintegrated parton distribution functions (uPDF's) and
gauge-invariant amplitudes with off-shell initial-state partons. The
second one is constructed in the same manner as it was suggested by
Kimber, Martin, Ryskin and Watt \cite{KMR,WMR}, but with sufficient
revision, see Ref.~\cite{NefedovSaleev2020}. The off-shell
amplitudes are derived using the Lipatov Effective Field Theory
(EFT) of Reggeized gluons~\cite{Lipatov95} and Reggeized
quarks~\cite{LipatovVyazovsky}. More details of PRA can be found in
Ref.~\cite{NSS2013}, the inclusion of real NLO corrections is
studied in Ref.~\cite{NKS2017}, the development of PRA in the full
one-loop NLO approximation is further discussed in
\cite{PRANLO1,PRANLO2,PRANLO3}.

Factorization formula of PRA for the process $p+p\to
\gamma\gamma\gamma + X$, can be presented in a $k_T$-factorized
form:
  \begin{eqnarray}
  d\sigma &=& \sum_{i,\bar j}\int\limits_0^1 \frac{dx_1}{x_1} \int \frac{d^2{\bf q}_{T1}}{\pi} {\Phi}_i(x_1,t_1,\mu^2)
\int\limits_0^1 \frac{dx_2}{x_2} \int \frac{d^2{\bf q}_{T2}}{\pi}
{\Phi}_{j}(x_2,t_2,\mu^2)\cdot d\hat{\sigma}_{\rm PRA},
\label{eqI:kT_fact}
  \end{eqnarray}
where $t_{1,2}=-{\bf q}_{T1,2}^2$, the off-shell partonic
cross-section $\hat\sigma_{\rm PRA}$ is determined by squared
Reggeized amplitude, $\overline{|{\cal A}_{PRA}|^2}$. Despite the
fact that four-momenta of partons in the initial state  are
off-shell ($q_{1,2}^2=-t_{1,2}<0$), the PRA hard-scattering
amplitude is gauge-invariant.
\subsection{New unintegrated PDFs}
To resolve collinear divergence problem, we require that uPDF
${\Phi}_i(x,t,\mu)$ in $(\ref{eqI:kT_fact})$ should be satisfied
exact normalization condition:
\begin{equation}
\int\limits_0^{\mu^2} dt \Phi_i(x,t,\mu^2) = {F}_i(x,\mu^2) \mbox{
or } \Phi_i(x,t,\mu^2)=\frac{d}{dt}\left[
T_i(t,\mu^2,x){F}_i(x,t)\right], \label{eq:sudakov}
\end{equation}
where $T_i(t,\mu^2,x)$  is referred as Sudakov form-factor,
satisfying the boundary conditions
$$T_i(t=0,\mu^2,x) = 0 \mbox{ and } T_i(t=\mu^2,\mu^2,x) = 1.$$ UPDF can be written as follows
from KMR model:
\begin{equation}
\Phi_i(x,t,\mu)= \frac{\alpha_s(\mu)}{2\pi} \frac{T_i(t,\mu^2,x)}{t}
\sum\limits_{j=q,\bar{q},g}\int\limits_x^1 dz\ P_{ij}(z) {F}_j\left(
\frac{x}{z}, t \right) \theta\left( \Delta(t,\mu)-z
\right).\label{uPDF}
\end{equation}
Here, we resolved also infra-red divergence taking into account the
cutoff:  $z < 1-\Delta_{KMR}(t_{1,2},\mu^2),$ where
$\Delta_{KMR}(t,\mu^2)=\sqrt{t}/(\sqrt{\mu^2}+\sqrt{t})$ is the
KMR-cutoff function~\cite{KMR}.

The solution for Sudakov form-factor in Eq. ({\ref{eq:sudakov}) has
been obtained in Ref.~\cite{NefedovSaleev2020} (see equations
(26)-(28)). There are important differences between the Sudakov
form-factor obtained in the PRA (\ref{uPDF}) and in the KMR
approach~ \cite{KMR}. At first, the Sudakov form-factor in PRA
contains the $x-$depended $\Delta \tau_i$-term in the exponent which
is needed to preserve exact normalization condition for arbitrary
$x$ and $\mu$. The second one is that in PRA the rapidity-ordering
condition is imposed both on quarks and gluons, while in KMR
approach it is imposed only on gluons.

\subsection{LO and NLO subprocesses}
In presented study, we take into consideration the LO subprocess of
three-photon production in quark-antiquark annihilation
\begin{equation}
Q \bar Q \to \gamma\gamma\gamma \label{LO}
\end{equation}
and NLO contributions of quark(antiquark)-gluon scattering
subprocesses
\begin{equation}
Q R \to q \gamma \gamma \gamma \label{NLO}.
\end{equation}
We don't consider NLO contributions from subprocesses $ Q\bar Q\to g
\gamma\gamma\gamma \label{QQ3gammaG}$
 which is negligibly small $(\leq 5 \%)$ at high energy.

In the Lipatov EFT, the LO (\ref{LO}) and NLO (\ref{NLO})
subprocesses are described by gauge-invariant sets of 28 and 238
Feynman diagrams, respectively. The direct integration of squared
amplitudes in the LO approximation of Lipatov EFT can be done in the
numerically-stable form. To calculate contributions from $2 \to 4$
NLO subprocesses with initial Reggeizaed partons such method is not
efficient and we use parton-level event generator KaTie
\cite{katie,hameren1,hameren2}. The LO contribution of subprocess
(\ref{LO}) were calculated for crosscheck  both with  event
generator KaTie \cite{katie} and semi-analytically with the help of
Feynman rules of Lipatov EFT. The approach used KaTie  for numerical
generation of off-shell amplitudes is equivalent to the Lipatov EFT
at the tree-level \cite{photon-jet,kutak}.

Next important usage is matching LO and NLO calculation in PRA
\cite{two-photon,NKS2017}. To extract specific double counting
between LO (\ref{LO}) and NLO (\ref{NLO}) subprocesses with emission
of additional quark which is separated in rapidity from three-photon
cluster. Such additional quark should be considered as emitted
parton during perturbative QCD evolution and  should be absorbed in
uPDF. Accordingly to KMR-PRA model of uPDFs, it has strong angular
(rapidity) ordering for emitted partons, such a way we should
extract events with rapidity configuration of final photons and
quark when rapidity of final quark smaller or large of photon's
rapidities, depending of sign of initial quark rapidity. Such
procedure decreases NLO contribution about 40-50 \%.

\vspace{-2cm}
\begin{figure}[!h]
\centering
\includegraphics[width=0.5\textwidth]{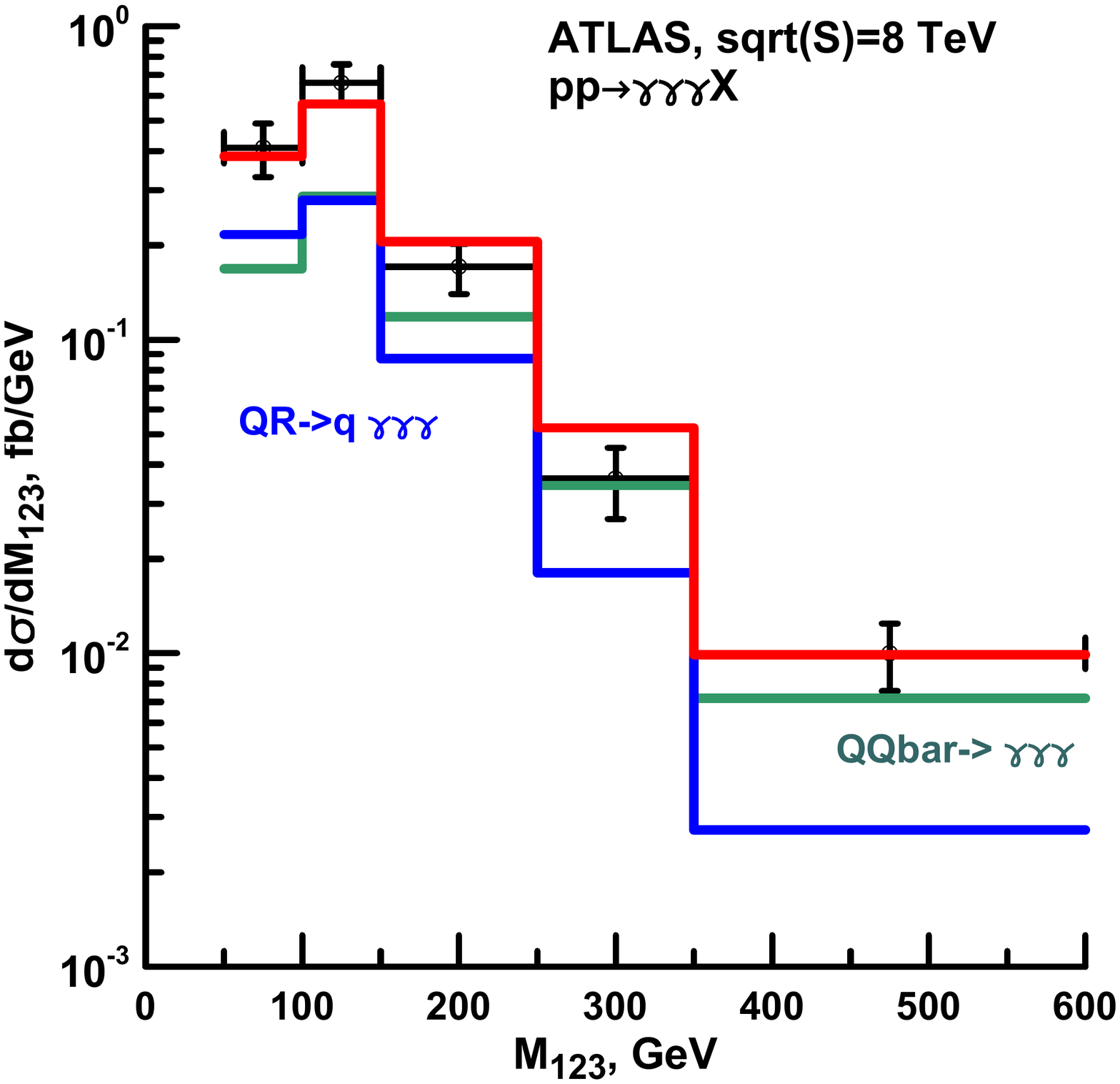}
\vspace{-2cm} \caption{The differential cross sections as function
of three-photon invariant mass $M_{123}$. The hard scale in PRA
calculation is taken as $\mu=M_{123}$. The green histogram
corresponds LO contribution (\ref{LO}), the blue histogram
corresponds NLO contribution (\ref{NLO}) and the red histogram is
their sum.} \label{Plot1}
\end{figure}

\section{Results}
\label{sec:res}
 We calculate cross section at different
choice of factorization ($\mu_F$)and renormalization ($\mu_R$)
scales, which we take equal to each other, $\mu_F=\mu_R=\mu$. In the
Table~$1$ we compare predictions obtained in PRA with
$\mu=M_{3\gamma}$ - invariant mass of three-photon system;
$\mu=k_{T,3\gamma}$ - sum of transverse momentum moduli; and
$\mu=E_{T,3\gamma}$ - transverse energy of three-photon system.
Table $2$ collects total cross sections at three energies
$\sqrt{s}=8$, $13$, and $27$ TeV. We compare PRA predictions with
result of calculation in NNLO CPM \cite{NNLO31,NNLO32}. PRA results
in LO with real NLO corrections are roughly coincide with full NNLO
predictions of CPM for $\sqrt{s}=8$ TeV. At higher energies (13 and
27 TeV) PRA predicts larger cross sections, up to $\sim 10$ \% and
$\sim 20$ \%, respectively.

The differential cross sections as function of three-photon
invariant mass $M_{123}$ is shown in Fig.~$1$. The hard scale is
taken as $\mu=M_{123}$. We find good agreement also for invariant
mass spectra of different photon-pairs $(M_{ij})$, rapidity
$(|\Delta y_{ij}|)$ and azimuthal angle $(|\Delta \phi_{ij}|)$
differences and transverse momenta of leading $p_{T1}$ and
subleading $(p_{T,2,3})$ photons.

\section{Conclusion}
We describe cross section and spectra for three-photon production in
LO PRA with real NLO corrections.  We demonstrate applicability of
new KMR-type uPDFs for using in HEF calculations.

\section*{Acknowledgements}
We are grateful to Maxim Nefedov for discussions and critical
remarks, and to Andreas van Hameren for helpful communications on MC
generator  KaTie.


\paragraph{Funding information}
The work has been supported in parts by the Ministry of Science and
Higher Education of Russia via State assignment to
   educational and research institutions under project FSSS-2020-0014.





 \bibliographystyle{SciPost_bibstyle} 

\nolinenumbers

\end{document}